\documentclass[10pt]{article}
             
\usepackage{graphicx}
\usepackage{mathrsfs}
\usepackage{latexsym}
\usepackage{amsmath}
\usepackage{amssymb}
\usepackage{color}
\usepackage{subfig}
\usepackage{array}
\usepackage{multirow}
\usepackage{hhline}
\usepackage[table]{xcolor}
\usepackage{authblk}
\usepackage{float}
\usepackage{subfig}

\usepackage[nohead,letterpaper]{geometry}
\usepackage{vmargin}
\setpapersize{USletter}
\setmargrb{2cm}{1cm}{2cm}{1.75cm}

\newenvironment{remark}[1][Remark]{\begin{trivlist}
\item[\hskip \labelsep {\bfseries #1}]}{\end{trivlist}}

\newcommand{\rmd}{\textrm{d}}

\usepackage{xcolor}
\definecolor{plum}{rgb}{0.36078, 0.20784, 0.4}
\definecolor{chameleon}{rgb}{0.30588, 0.60392, 0.023529}
\definecolor{cornflower}{rgb}{0.12549, 0.29020, 0.52941}
\definecolor{scarlet}{rgb}{0.8, 0, 0}
\definecolor{brick}{rgb}{0.64314, 0, 0}

\usepackage[%
  pdftitle={},%
  pdfauthor={Miok Park},%
  pdfsubject={},%
  pdfkeywords={},%
  colorlinks=true,linkcolor=cornflower,citecolor=brick,urlcolor=chameleon%
  ]{hyperref}

\begin{document}



\begin{flushright}
 CTPU-PTC-23-42
\end{flushright}
~\vspace{1cm}\\

\begin{center}
{\bf \Large Thermodynamics with conformal Killing vector in the charged Vaidya metric}
\end{center}

\vspace{.5cm}

\begin{center}
Seoktae Koh$^{a,b}$, Miok Park$^{c}$, Abbas M. Sherif$^{a,} \footnote{All authors contributed equally to this work.}$

\vspace{0.5cm}{\small{a. Department of Science Education, Jeju National University, Jeju, 63243, South Korea} \\
\small{b. Institute for Gravitation and the Cosmos, Pennsylvania State University, University Park, PA 16802, USA}\\
\small{c. Particle Theory  and Cosmology Group, Center for Theoretical Physics of the Universe, Institute for Basic Science (IBS), Daejeon, 34126, Korea}}
\end{center}
\vspace{0.5cm}

\begin{center}
{\bf Abstract}
\end{center}

We investigate the charged Vaidya spacetime with conformal symmetry by classifying the horizons and finding its connection to Hawking temperature. We find a conformal Killing vector whose existence requires the mass and electric charge functions to be proportional, as well as linear in time. Solving the Killing equations for the conformally transformed metric from the linear charged Vaidya metric yields the required form of the conformal factor. From the vanishing of the norm of the conformal Killing vector, we find three conformal Killing horizons which, under the transformation, are mapped to the Killing horizons of the associated static spacetime, if the spherical symmetry is maintained. We find that the conformal factor is not uniquely determined, but can take any function of the ratio of the radial coordinate to the dynamical mass. As an example, we illustrate a static spacetime with our choice of the conformal factor and explicitly show that the surface gravity of the conformal Killing horizons, which is conformally invariant, yield the expected Hawking temperature in the static spacetime. This static black hole spacetime contains a cosmological horizon, but it is not asymptotically de Sitter. We also investigate the case when the mass parameter is equal to the constant electric charge. While in this case the standard pair of horizons, the loci of the time component of the metric, degenerate, the conformal Killing horizons do not degenerate. This therefore leads to a non-zero Hawking temperature in the associated static spacetime.

\newpage

\setcounter{page}{1}


\tableofcontents


\newpage

\section{Introduction}

Describing evaporating or collapsing black holes is essential for studying the astrophysical and quantum properties of black holes. However, finding mathematical or numerical solutions for these systems is challenging. Vaidya spacetime is one of the rare examples that can describe such dynamical systems. Its analytic form allows us to understand some properties of dynamical black hole spacetimes, which requires distinct notions from that of static or stationary black holes such as Schwarzschild and Kerr-Newman solutions in asymptotically flat spacetimes. For example, the event horizon is no longer applicable to characterize dynamical black holes since the notion is a global property of the spacetime, defined as the causal boundary beyond which events cannot affect an observer. Since dynamical spacetimes change in time, this global notion of horizons is not well defined. Instead, we must use (quasi-)local concepts of horizons. To this end, there have been studies to construct more general characterizations to capture local properties of black holes.

In a seminal work by Roger Penrose in \cite{Penrose:1964wq}, for which he was awarded the Nobel Prize in Physics in 2020, it was proved that the formation of a singularity is an inevitable consequence of gravitational collapse, under mild assumptions on the energy conditions on the spacetime. In his proof, he introduced the concept of a trapped surface, which is defined by the negativity of the expansions of both outgoing and ingoing null geodesics to surfaces. The very first local characterizon of horizons was initiated by Hayward \cite{Hayward:1993wb}, where he introduced \textit{trapping horizons}, foliated by marginaly outer trapped surfaces on which the the expansion of the outgoing null geodesics vanish. The idea is that under certain conditions, just to the inside of these horizons, there will be trapped surfaces. Since then, several detailed classifications have been constructed (see \cite{Ashtekar:2004cn,Ashtekar:2005ez}, where the notions of \textit{isolated} and \textit{dynamical} horizons were introduced). However, all of these notions are not yet consistent in the literature (see \cite{Booth:2005bound,Booth:2005ng} for an excellent reviews of the topic). For this reason, when classifying horizons, we will specify the convention that we use in this paper in Section 2.

Compared to the interest in evolving spacetime models, spacetimes with conformal symmetry have not yet been observed for astrophysical objects. However, this symmetry is often observed in laboratories, such as at quantum critical points and some fixed points of RG flow in condensed matter systems \cite{PhysRevB.23.4615,PhysRevLett.35.1678,sachdev_2011}. 
Therefore, we should not completely exclude the existence of such objects in our universe. 

Conformal symmetries in spacetimes are represented by conformal Killing vectors, infinitesimal generators of the symmetries, which are solutions to the conformal Killing equations. The existence of these vectors generates new solutions via a conformal transformation. Provided that (irrotational) observers of conformal motion can be defined, where the norm of a conformal Killing vector vanishes, a limit surface is formed, known as a conformal Killing horizon \cite{dyer:1979a}. This horizon maps to the Killing horizon of the associated static spacetime. The notion of surface gravity is defined for conformal Killing horizons as well as Killing horizons. Surface gravity can be defined in three different ways for conformal Killing vectors, as discussed in \cite{Jacobson:1993pf}, which for (true) Killing vectors, these three definitions all agree. The relationship between surface gravity associated with conformal Killing vectors and thermodynamics has been studied in, for example, \cite{Jacobson:1993pf,Nielsen:2017hxt}.

In asymptotically flat spacetime, black hole solutions possess at least two Killing vectors. However, a black hole solution with only one Killing vector was later found \cite{Dias:2011at,Stotyn:2011ns}, which has scalar hairs. This suggests that dynamical black hole solutions may have less symmetry or no isometries at all. However, if we restrict ourselves to the spherically symmetric case, dynamical spacetimes can have scaling symmetry, which is associated with conformal Killing vectors. The linear Vaidya spacetime is an example of this, where by ``linear" it is meant that the mass function is linear in time. Recent studies examined conformal symmetries in the uncharged linear Vaidya spacetime which is asymptotically flat \cite{galaxies2010062,Nielsen:2017hxt}. (For the case of more general mass function, where the mass depends on both the radial and temporal coordinates, see the references \cite{ojako2020conformal,vertogradov2023generalized} and those therein.) If the time-dependent variables in the metric are bounded, then the Vaidya spacetime can be considered asymptotically flat. This is interesting because asymptotically flat and static (or stationary) spacetimes has Spi group at spatial infinity and BMS group at null infinity, which are larger than Poincaré group but do not contain conformal symmetry. In contrast, there are spacetimes that intrinsically contain conformal group, such as (anti) de Sitter spacetime, which played an important role in constructing the AdS/CFT correspondence \cite{Maldacena:1997re,Witten:1998qj,Gubser:1998bc}.

Motivated by the linear Vaidya spacetime, we investigate the charged Vaidya spacetime with conformal symmetry by classifying the horizons and finding its connection to the Hawking temperature. First, we find the conformal Killing vectors for the given spacetime. Their existence requires the mass and electric charge functions to be proportional and linear in time. We express the linear charged Vaidya solution in the ingoing Eddington-Finkelstein coordinate. By solving the Killing equations for the conformally transformed linear charged Vaidya metric, we obtain the conformal factor $\Omega$. Thus, the conformal transformation automatically maps the conformal Killing horizons in the dynamical spacetime to the Killing horizons in the static spacetime, if the spherical symmetry is maintained. We find that the conformal factor is not uniquely determined, but can take any form of a function of $r/m(v)$. As an example, we illustrate a static spacetime with our choice of $\Omega$ and explicitly show that the surface gravity, which is conformally invariant and associated with the conformal Killing horizon, yields the Hawking temperature in the static spacetime. We also show that when the mass parameter $M$ is equal to the constant electric charge $Q$, the conformal Killing horizons do not degenerate. This is unlike the case of the horizons located at vanishing points of the $(v,v)-$component of the metric, which become degenerate. This therefore leads to the conclusion that the Hawking temperature in the static spacetime is non-zero. 

We organize the paper as follows: In section 2, we derive the solutions to the Einstein and Maxwell equations that are compatible with the conformal Killing equations. We then explore the corresponding static spacetimes via conformal mapping. In section 3, we analyze the horizons generated by the dynamical spacetime, and compute the surface gravity at the conformal Killing horizons. We find that the conformal Killing horizons in the dynamical spacetime map to Killing horizons in the static spacetime, so that the covariantly invariant surface gravity is identified as the Hawking temperature in the static spacetime. We also consider a special case where the mass is equal to the electric charge. For RN black holes, for example, this leads to a degeneracy of the Killing horizons, which in turn yields a zero Hawking temperature for RN black holes. We show that while the standard pair of horizons of the linear charged Vaidya spacetime are degenerate, but the conformal Killing horizons are not. Therefore, the surface gravity takes non-zero values in the dynamical spacetime, and so it gives a non-zero Hawking temperature in the associated static spacetime. We conclude in section 4.

\section{Charged Vaidya metric}

For the purpose of this work, we are interested in dynamical spacetimes admitting a black hole. One simple and well-known example is the Vaidya spacetime with infalling or outgoing null dust. The conformal Killing horizons have been found and studied in the uncharged case \cite{Nielsen:2017hxt}. Our interest here is to extend this study to the Vaidya spacetime with electric charge \cite{Vaidya:1970v1}.  In this section we introduce the metric of interest, and analyze the conformal Killing equations. We then consider the static spacetime that is conformally mapped to this charged metric.

\subsection{The metric}

The charged Vaidya metric in the ingoing Eddington-Finkelstein coordinates is written as
\begin{align}\label{gvc1}
ds^2=-f(v,r)dv^2+2dvdr+r^2d\Omega_2,
\end{align}
where the metric function takes a form
\begin{align}\label{can0}
f(v,r) = 1 - \frac{2m(v)}{r} + \frac{q(v)^2}{r^2}, 
\end{align}
and $d\Omega_2$ denotes the two-sphere line element. The electric potential is taken as
\begin{align}
A = \frac{1}{\sqrt{4 \pi}} \bigg(\frac{q(v)}{r} -\frac{q(v)}{r_0} \bigg) \rmd v
\end{align}
where $r_0$ is a gauge freedom that can be chosen to be any value. These solutions describe a spacetimes that absorbs matter fields. To satisfy Einstein and Maxwell equations
\begin{align}
R_{ab} - \frac{1}{2} R g_{ab} = 8 \pi G T_{ab} = 8 \pi G (T^{(\textrm{EM})}_{ab}+T^{(\textrm{ex})}_{ab}), \qquad \nabla^{a} F_{ab} = j_{b},
\end{align}
the energy-momentum tensor and the current are required to be
\begin{align}
{T^{(\textrm{EM})}}^{a}{}_{b} = \frac{q(v)^2}{8 \pi r^4} \textrm{diag}(-1,-1,1,1), \qquad  T^{(\textrm{ex})}_{vv} = \frac{2}{8 \pi r^3} \left(r \dot{m}(v)-q(v) \dot{q}(v)\right), \qquad j_v = - \frac{\dot{q}(v)}{2 \sqrt{\pi } r^2}
\end{align}
where $T^{(\textrm{EM})}_{ab}$ and $T^{(\textrm{ex})}_{ab}$ are the contribution from the electromagnetic field and the null dust respectively. Since $T^{(\textrm{EM})}_{ab} k^{a} k^{b}=0$, in order to satisfy the null energy condition ($T_{ab} k^a k^b \ge 0$, where $k^a$ is any null vector) and all other standard energy conditions, the following condition is demanded: 
\begin{align}
r \dot{m}(v)-q(v) \dot{q}(v) \geq 0,\label{eq:ECdn}
\end{align}
where the left hand side is just the energy density of the null dust (see the reference \cite{Booth:2015kxa} for more discussion). The spacetime always satisfy this condition when $q(v)=0$, but there is a chance to violate this condition when $q(v) \neq 0$. More details on this energy condition will be discussed in section 3.2.

\subsection{Conformal Killing vector for the charged Vaidya metric}

We here examine if the charged Vaidya metric is invariant under conformal transformation. 

\subsubsection{Conformal transformations}

A spacetime $M$ possesses a conformal Killing vector (CKV, henceforth) $\xi^a$ provided that the Lie derivative of the spacetime metric along $\xi^a$ scales the metric:
\begin{align}\label{cke}
\mathcal{L}_{\xi}g_{ab}=2\varphi g_{ab}. 
\end{align}
This is called the conformal Killing equation (CKE), where $\varphi$ is a smooth function on $M$ defined by $4\varphi=\nabla_a\xi^a$. If $\varphi$ is zero, constant, or non-constant, $\xi^a$ is a true Killing vector (KV), a homothetic Killing vector (HKV), or a (proper) conformal Killing vector (CKV) respectively. 

The existence of a conformal Killing vector implies (and is implied by) the existence of a metric $\bar g_{ab}$, related to $g_{ab}$ as

\begin{align}\label{cr1}
g_{ab} = \Omega^2 \bar g_{ab}, \qquad g^{ab} = \Omega^{-2} \bar g^{ab}, \qquad \Omega \neq 0,
\end{align}
for smooth $\Omega$. 

Now, suppose the spacetime $\bar M$ with metric $\bar g_{ab}$, is static, i.e. admits a globally timelike KV $\bar\xi^a$, so that

\begin{align}
\mathcal{L}_{\bar{\xi}}\bar g_{ab}=0.
\end{align}
From \eqref{cr1} it is checked that $\xi^a$ is a CKV for the spacetime with conformal metric $g_{ab}$ if and only if $\Omega$ is non-constant, in which case $\Omega$ verifies
\begin{align}\label{osol}
\left(\mathcal{L}_{\xi}-\varphi\right)\Omega&=0.
\end{align}
The vectors in the static and conformal spacetimes are related as
\begin{align}
\xi^a=\bar\xi^a;\quad \xi_a= g_{ab}\xi^b=\Omega^2\bar g_{ab}\xi^b=\Omega^2\bar\xi_a \label{eq:CKnK}.
\end{align}
If the static spacetime $\bar M$ is a black hole spacetime with a Killing horizon $\bar{\mathcal{T}}$ generated by $\bar\xi^a$, the conformal spacetime has a black hole with a conformal Killing horizon $\mathcal{T}$, and vice versa, by virtue of the invariance of the causal structure under the conformal map.

\subsubsection{The conformal Killing vector}

To find a CKV for the metric \eqref{gvc1}, we solve the CKE \eqref{cke}. To this end, the following vector ansatz is employed:
\begin{align}
\xi^{a} = \{h(v,r), \bar{h}(v,r), 0,0 \}, \qquad \xi_a = \{ -f(v,r) h(v,r) +\bar{h}(v,r), h, 0, 0 \}.
\end{align}
The CKE then expands, component-wise, to the following set of equations:
\begin{align}
0& = h'(v,r), \label{eq:CKEvv} \\
0&  =  \frac{2 \bar{h}(v,r)}{r} + 2 f(v,r) h'(v,r)+ \bar{h}'(v,r) + \dot{h}(v,r), \\
0& = - \bar{h}(v,r)f'(v,r) - h(v,r) \dot{f}(v,r) + f(v,r) \left(-\frac{\bar{h}(v,r)}{r} + \frac{3}{2} \bar{h}'(v,r) - \frac{1}{2} \dot{h}(v,r) \right) + 2 \dot{\bar{h}}(v,r), \label{third}\\
0& = 2 \bar{h}(v,r) -r (\bar{h}'(v,r) + \dot{h}(v,r)), 
\end{align}
where the prime($'$) notates the variation with respect to $r$. Solving the first and second (or fourth) equations give $h(v,r)$ and $\bar h(v,r)$ respectively as

\begin{align}
h(v,r) = h(v),  \qquad \bar{h}(v,r) = c_1(v)  r^2 + r \dot{h} (v),
\end{align}
where $c_1(v)$ is the integration constant. Substituting these to \eqref{third}, we obtain the following second order ODE in $h(v)$
\begin{align}
\ddot{h}(v) + c_1(v) f(v,r) - \frac{1}{2} \bigg( r c_1(v) + \dot{h}(v) \bigg) f'(v,r)- \frac{1}{2 r} h(v) \dot{f}(v,r) = 0. \label{eq:finCKE}
\end{align}
Plugging the metric function $f(v,r)$ from \eqref{can0} yields
\begin{align}
0&=\ddot{h}(v)+\frac{q(v)}{r^3}\left(q(v) \dot{h}(v) -h(v) \dot{q}(v) \right) + \frac{1}{r^2} \left(2 q(v)^2 c_1(v)-m(v) \dot{h}(v)+h(v) \dot{m}(v)\right)\nonumber\\
&-\frac{3 m(v) c_1(v)}{r}+r \dot{c_1}(v) + c_1(v) .
\end{align}
Since all variables only depend on $v$, each term in the same order of $r$ should vanish. The first term in the second line indicates $c_1(v)=0$. Therefore, it is required that
\begin{align}
0&=\ddot{h}(v),  \label{eq:CKE1} \\
0& = q(v) \dot{h}(v) -h(v) \dot{q}(v), \label{eq:CKE2} \\
0&  = -m(v) \dot{h}(v)+h(v) \dot{m}(v), \label{eq:CKE3} 
\end{align}
The equations (\ref{eq:CKE2}) and (\ref{eq:CKE3}) have the same structure for the interchange of $m(v)$ and $q(v)$, yielding the solutions 
\begin{align}
h(v)& = c_2 m(v),  \; \; \textrm{or} \; \; h(v) = c_3 q(v), \label{eqg3}
\end{align}
where $c_2$ and $c_3$ are integration constants. The result \eqref{eqg3} implies that the mass function $m(v)$ should be proportional to the electric charge function $q(v)$, say $m(v) = \gamma q(v)$, for some constant $\gamma$. The first condition (\ref{eq:CKE1}) implies that both the mass and the electric charge functions can be at most of linear order in $v$: 
\begin{align}
m(v) = M + \mu (v-v_0), \qquad q(v) = \frac{Q}{M} m(v) = Q + \frac{\mu Q}{M} (v-v_0),\label{eq:mandq}
\end{align}
where $\mu$ is interpreted as the mass density of the null dust. We have fixed $\gamma = Q/M$ so as to recover Reissner–Nordström black holes in the limit \(v\to v_0\) or $\mu \rightarrow 0$. Then, the form of the metric function $f(v,r)$ satisfying the CKE is necessarily
\begin{align}
f(v,r) = 1 - \frac{2m(v)}{r} + \frac{Q^2}{M^2} \frac{m(v)^2}{r^2}.  \label{eq:CKmetric}
\end{align}
We therefore refer to this Vaidya solution as the \textit{linear} charged Vaidya (LCV) metric.

Finally, choosing the integration constant $c_2=1/M$ gives the following solution to the CKE:
\begin{align}
\xi^{a} = \bigg\{ \frac{m(v)}{M}, \frac{r \mu}{M}, 0,0 \bigg \}.\label{cckv}
\end{align}
This choice of the constant $c_2$ normalizes the CKV to unity at infinity in the static limit \(\mu\to 0\). 

\begin{remark}
The CKV \eqref{cckv} found here has the exact same form as that found by Nielsen \cite{Nielsen:2017hxt} for the uncharged case (of course, its dual contains $q(v)$ due to the presence of the metric function $f(v,r)$). The choice $\ddot{h}=0=c_1$ appears to be the only configuration which produces a solution to \eqref{eq:finCKE}. For example, if \(\ddot{h}\neq0\) or \(c_1\) a non-zero constant, the equation \eqref{eq:finCKE} appears intractible. Moreover, for the constant charge case, we find that there are no CKVs. While we do not present an explicit proof, we intuit that the form \eqref{cckv} appears to be the only solution to the CKE for the charged Vaidya metric. Whereas, in a recent paper, \cite{tarafdar2023slowly}, the authors introduced the ansatz $\xi^a=\{v,r,0,0\}$, and found that this satisfies the homothetic Killing equation with $\varphi=1$, provided the mass and charge obey $m(v)=v\dot{m}(v)$ and $q(v)=v\dot{q}(v)$. Here we have approached this systematically by starting with a general spherically symmetric ansatz for $\xi^a$, and then solving the conformal Killing equations. Having a solution to the conformal Killing equations then conditions the mass and charge to obey the general linear forms \eqref{eq:mandq}.
\end{remark}

\subsection{Conformal mapping to static spacetimes}

We now explore the static spacetime $\bar{g}_{ab}$, which is conformally mapped from the charged Vaidya spacetime $g_{ab}$. Firstly, computing $\varphi$ from (\ref{cckv}) we obtain
	\begin{align}
	\varphi= \frac{1}{4} \nabla_{a} \xi^{a} = \frac{\mu}{M},
	\end{align}
suggesting that $\xi^a$ is in fact a homothetic Killing vector and not a proper CKV, as is the case with the CKV in \cite{Nielsen:2017hxt}. To obtain the associated conformal factor $\Omega^2(v,r)$, we solve the differential equation \eqref{osol} to obtain
	\begin{align}
	\Omega^2(v,r)= \bigg(\frac{m(v)}{M}c_4(v,r) \bigg)^2.
	\end{align}
Ensuring that $\xi^a$ is a Killing vector for the associated static metric $\bar g_{ab}$ necessitates that the function $c_4(v,r)$ takes the form $c_4(v,r) = c_4 (r/m(v))$, with no other constraints. This suggests that the conformal factor  $\Omega(v,r)$ is not uniquely determined. Rather, each choice of $c_4(v,r)$ leads to a different static spacetime $\bar{g}_{ab}$ from the same dynamical spacetime $g_{ab}$.

Let us introduce a new variable $y$ as
\begin{align}
y = M \frac{r}{m(v)},
\end{align}
so that 
\begin{align}
c_4\left(\frac{r}{m(v)} \right) = c_4(y), \qquad F(v,r) = F(y) = 1 - \frac{2 M }{y} + \frac{Q^2}{y^2},
\end{align}
and then redefine \(dv=(m(v)/M)dV\). The conformal mapping in (\ref{cr1}) then induces a static metric $\bar{g}_{ab}$ from $g_{ab}$. The new metric $\bar g_{ab}$ can be expressed in terms of the new variables $V$ and $y$ as follows:
\begin{align}
\rmd \bar{s}^2 = -\frac{M F(y) - 2 \mu  y}{M c^2_4(y)} \rmd V^2 + \frac{2}{c^2_4(y)} \rmd V \rmd y + \frac{y^2}{c^2_4(y)}  (\rmd \theta^2 + \sin^2 \theta \rmd \phi^2). \label{eq:conformalmap}
\end{align}
We further make a coordinate transformation
\begin{align}
R = \frac{y}{c_4(y)},
\end{align}
where $c_4(y)$ can be any function of $y$, but cannot be proportional to $y$. Then the static metric takes the following form:
\begin{align}
\rmd \bar{s}^2 =-\frac{M F(y) - 2 \mu  y}{M c^2_4(y)} \rmd V^2 + \frac{2}{(c_4(y) - y c_4'(y))} \rmd R \rmd V + R^2 \rmd \Omega_2.
\end{align}
Once $c_4$ is specified, $y$ can be expressed in terms of $R$. The simplest case that one could consider is for $c_4(y)$ to be a constant. If we set $c_4(y) = 1$, then $R = y$. The static metric can then be written in the following form:
\begin{align}
\rmd \bar{s}^2 = - \tilde{F}(R)\rmd V^2 + 2 \rmd R \rmd V + R^2 \rmd \Omega_2,\label{eq:metricC21}
\end{align}
where
\begin{align}
\tilde{F}(R) = 1 - \frac{2 M }{R} + \frac{Q^2}{R^2} - \frac{2 \mu}{M} R.
\end{align}
This spacetime is neither asymptotically flat nor asymptotically de Sitter. This spacetime appears as a solution obtained in \cite{Kiselev:2002dx}, which is a black hole solution surrounded by the quintessence matter. It also appears as the solution obtained when a Maxwell field and anisotropic matter are employed in gravity action \cite{Jeong:2023hom}. 

We could also consider the following form of the function $c_4(y)$ 
\begin{align}
c_4(y) = \sqrt{\frac{2\mu}{M} y}. 
\end{align}
This is a choice that makes the norm of Killing vectors to be normalized at infinity. The corresponding static metric in this case is expressed as
\begin{align}
\rmd \bar{s}^2 = \left(1 -\frac{M^2}{4 \mu ^2 R^2} +\frac{M^4}{4 \mu ^3 R^4} - \frac{M^4 Q^2}{16 \mu ^4 R^6} \right) \rmd V^2 + \frac{M}{R \mu} \rmd V \rmd R + R^2 \rmd \Omega_2.
\end{align}
However, as is seen, this metric is not asymptotically flat. This implies that the asymptotic flatness of the linear charged Vaidya spacetime with bounded mass function, is not recovered in the the corresponding static spacetime under the conformal map.

\section{Horizons in the linear charged Vaidya metric}

The metric \eqref{gvc1} with function (\ref{can0}) has horizons located at radii $r_{\pm}$ where the metric function $f(v,r)$ vanishes. In addition to these horizons, there are conformal Killing horizons where the norm of the CKV (\ref{cckv}) vanishes. These two kinds of the horizons do not coincide. It has been argued that \cite{Nielsen:2017hxt} conformal Killing horizons are the natural objects to associate to black holes in dynamical spacetimes whenever we have a conformal map to a static black hole spacetime. In such case, of course, one expects the outer conformal Killing horizon to be located `outside' the $f(v,r_{+})=0$ horizon. We briefly investigate the properties of the $f(v,r_{\pm})=0$ horizons in the LCV metric, and classify them using the notion of marginally outer trapped surfaces (we follow some standard literature on the subject, see for example \cite{Booth:2005bound}). We then compare aspects of these horizons to the conformal Killing ones. We compute surface gravity at each conformal Killing horizon and associate these surface gravities with temperature. 


\subsection{Marginally outer trapped surfaces}

For a given spacetime $M$ and an embedded $2$-surface \(\Sigma\subset M\), we can construct a pair of null normal vectors $l^a$ and $n^a$ to $\Sigma$, satisfying $l^a l_a = n^{a}n_a = 0$ and $l^a n_a = - 1$, where we take $l^a$ and $n^a$ to denote, respectively, the outgoing and ingoing null vectors to $\Sigma$. We denote the intrinsic metric to the $2$-surface $\Sigma$ as $\sigma_{AB}$, the pull-back of 

\begin{align}
\sigma_{ab} = g_{ab} + l_{a}n_{b} + n_{a} l_{b},
\end{align}
to $\Sigma$, where the indices $A$ and $B$ label the coordinates on $\Sigma$.

The expansions of the outgoing and ingoing null normal vectors $l^{a}$ and $n^{a}$ are given by
\begin{align}
\chi = \sigma^{ab}\nabla_al_b,\quad \tilde{\chi} = \sigma^{ab}\nabla_an_b.
\end{align}
A marginally outer trapped surface (MOTS) is a surface on which the expansion \(\chi\) vanishes. Within the context of black holes, it is required that the expansion \(\tilde{\chi}\) is strictly negative. In this case, `MOTS' is abbreviated to `MTS' (marginally trapped surface). Surface are additionally classified, based on the signs of these functions as follows: 
\begin{itemize}	
	\item Untrapped surfaces : $\chi > 0, \; \; \tilde{\chi} < 0$;
	\item Trapped surfaces : $\chi < 0, \; \;  \tilde{\chi} < 0$.
\end{itemize}

A 3-dimensional hypersurface foliated by MOTS is referred to as a marginally outer trapped tube (MOTT) \cite{Booth:2005ng}. And if the foliation is by MTS, the MOTT is simply a marginally trapped tube (MTT). On a MTT, there always exists a function $C$ such that the vector $X^a=l^a-Cn^a$ is tangent to the MTT (in which case the normal vector to the MTT is $\bar X^a=l^a+Cn^a$). Indeed,  $X_aX^a=2C$, so that the MTT is spacelike, timelike or null at points where $C>0,C<0,$ or $C=0$. The causal character of a MTT may vary at different points. In the case that the causal character is fixed across the MTT (this will be the case for spherical symmetry), the MTT is characterized as follows \cite{Booth:2005ng}: For $C>0,C<0,$ or $C=0$ the MTT is a dynamical horizon (DH), a timelike membrane (TLM), or an isolated horizon (IH). 

Explicitly, one can obtain $C$ from the relation
\begin{align}
C=\frac{\mathcal{L}_l\chi}{\mathcal{L}_n\chi}. \label{eq:C}
\end{align}
This follows from the fact that the expansion $\chi$ remains zero along the MTT, i.e. $\mathcal{L}_X\chi=0$.  From now on, we simply refer to an MTT as horizon.

For a horizon to enclose a trapped region, i.e. a black hole, it is required that $\mathcal{L}_n\chi<0$ \cite{Hayward:1993wb,Booth:2005ng}. (This condition is related to the notion of MOTS (MTS in our case) stability introduced by Andersson and Metzger \cite{Lars:2005stable}.)  If the null energy condition is imposed, $\mathcal{L}_{\ell}\chi\leq0$. Thus, the MTT is an isolated horizon if and only if the null energy condition is marginally satisfied, and a dynamical horizon only if $\mathcal{L}_{\ell}\chi<0$. 

It was recently discussed in \cite{Dunsby:2023stable}, in the context of perfect fluid locally rotationally symmetric (LRS) spacetimes, that the strict stability requires that not only both $\mathcal{L}_l \chi$ and $\mathcal{L}_n\chi$ are negative, but also that the scalar $C$ is sandwiched as $0<C<1$. This will be true in spherical symmetric, and can be most readily seen by writing the tangent to the horizon in terms of the unit timelike and spacelike directions orthogonal to the MTS. 

Now, explicitly the null vectors associated with the metric (\ref{gvc1}) are written as
\begin{align}
\mbox{Outgoing:}\qquad\left\{
\begin{array}{ll}
      l^a&=(1,\frac{f}{2},0,0) \\
      l_a&=(-\frac{f}{2},1,0,0),
\end{array} 
\right.\\
\mbox{Ingoing:}\qquad\left\{
\begin{array}{ll}
      n^a&=(0,-1,0,0) \\
      n_a&=(-1,0,0,0).
\end{array} 
\right. 
\end{align}
Constructing $\sigma^{ab}$ from the above, we can read off the pull-back $\sigma^{AB}$ from the 2-sphere metric
\begin{align}
\rmd s_{\Sigma}^2 = \sigma_{AB} \rmd x^{A} \rmd x^{B} = r^2 \left(\rmd \theta^2 + \sin^2 \theta \rmd \phi^2\right). 
\end{align}
The associated null expansions $\chi$ and $\tilde{\chi}$ are respectively
\begin{align}
\chi=\frac{f(v,r)}{r},\quad\tilde{\chi}=-\frac{2}{r}<0,
\end{align}
from which it follows that the MTS condition for the 2-spheres is $f(v,r)=0$, yielding the locations of the horizons as
\begin{eqnarray}
r_{\pm}=\frac{m(v)}{M}\left(M\pm\sqrt{M^2-Q^2}\right). \label{eq:rpm}
\end{eqnarray}
We refer $r_{+}$ as the outer horizon and $r_{-}$ as the inner horizon. (Clearly, it is required that $M \geq Q$) The Lie derivatives of $\chi$ along $l^a$ and $n^a$ are respectively
\begin{align}
\mathcal{L}_l\chi= l^{a} \partial_a \chi =\frac{f(v,r)f'(v,r)}{2 r}-\frac{f^2(v,r)}{2 r^2} + \frac{\dot f(v,r)}{r},\quad\mathcal{L}_n\chi = n^{a} \partial_a \chi =\frac{f(v,r)}{r^2} -\frac{f'(v,r)}{r},
\end{align}
which, upon evaluating on the MOTS gives
\begin{align}
\mathcal{L}_l\chi |_{r_{\pm}} &= \frac{\dot f(v,r)}{r} \bigg|_{r_{\pm}} =-\frac{2\mu m(v)}{M^2r_{\pm}^3}\left[\left(M^2-Q^2\right)\pm M\sqrt{M^2-Q^2}\right],\\
\mathcal{L}_n\chi |_{r_{\pm}} &= -\frac{f'(v,r)}{r} \bigg|_{r_{\pm}} = - \frac{2m(v)^2}{M^2r_{\pm}^4}\left[\left(M^2-Q^2\right)\pm M\sqrt{M^2-Q^2}\right].
\end{align}
The scalar $C$ then takes the conveniently simple form 
\begin{align}
C = \frac{\mu r_{\pm}}{m(v)} = \frac{\mu}{M}\left(M\pm\sqrt{M^2-Q^2}\right). \label{eq:C}
\end{align}
At $r=r_{+}$, it is clear that $\mathcal{L}_{\ell}\chi<0$ and $\mathcal{L}_n\chi<0$. This gives  $C>0$ and thus, the outer horizon is a dynamical horizon enclosing a black hole. At $r=r_{-}$, it becomes that $\chi = 0$, $\mathcal{L}_{\ell}\chi > 0$ and $\mathcal{L}_n\chi>0$. Therefore, $C$ is positive. While this implies that the inner horizon is a dynamical horizon, it cannot have trapped surfaces just to its `inside',  and the null energy condition fails there.

As was earlier mentioned, the stability condition of MTS requires that $C$ lies in the unit open interval $0<C<1$. Applying this condition to the scalar $C$ in (\ref{eq:C}), this imposes the following cut-off on $\mu$ 

\begin{align}
\mu<\frac{1}{1\pm\sqrt{1-Q^2/M^2}}.\label{eq:mubound}
\end{align}
For example, the MTS stability condition requires the cut-off on $\mu$ to be $\mu<1/2$ for pure Vaidya ($Q=0$). In the static limit $\mu\to0$, where one recovers the R-N case, \eqref{eq:mubound} always holds. The parameter $\mu$ is a substantially small parameter in astrophysics processes, see for example \cite{Nielsen:2010astro} where this parameter has an order of about $10^{-21}$, so that the bound \eqref{eq:mubound} always hold.

\subsection{Conformal Killing Horizons}

The conformal Killing vector obtained in (\ref{cckv}) generates conformal Killing horizons if its norm vanishes at a real value of $r$:
\begin{align}
\xi^{a} \xi_{a} = \frac{m(v) \left(-m(v) \left(q(v)^2+r^2\right)+2 r m(v)^2+2 \mu  r^3\right)}{M^2 r^2} = 0.
\end{align}
This cubic equation in $r$ yields the following solutions:
\begin{align}
r_1 &= \frac{m(v)}{6 \mu } \bigg(1+\sqrt[3]{b} + \frac{1-12 \mu }{\sqrt[3]{b}} \bigg), \\
r_2 &=\frac{m(v)}{6 \mu } \bigg(1-\frac{\left(1-i \sqrt{3}\right) \sqrt[3]{b}}{2}   -\frac{2 (1-12 \mu )}{\left(1-i \sqrt{3}\right) \sqrt[3]{b}}\bigg), \\
r_3 & =\frac{m(v)}{6 \mu } \bigg(1-\frac{\left(1+i \sqrt{3}\right) \sqrt[3]{b}}{2}   -\frac{2 (1-12 \mu )}{\left(1+i \sqrt{3}\right) \sqrt[3]{b}}\bigg),
\end{align}
where
\begin{align}
b=1 -18 \mu +\frac{6 \mu}{M^2} \left(9 \mu  Q^2 + i \sqrt{3} \sqrt{(1 - 16 \mu) M^4+(18 \mu -1) M^2 Q^2 - 27 \mu ^2 Q^4}\right).
\end{align}
We will refer $r_1, r_2$ and $r_3$ as cosmological conformal Killing horizon, inner conformal Killing horizon, and outer conformal Killing horizon respectively. These three roots take real distinct values if the expression inside the square root in $b$ is positive. This condition restricts the range of $\mu$ as follows:
\begin{align}
0 < \mu < \frac{M \left(-8 M^3+9 M Q^2 + \left(4 M^2-3 Q^2\right)^{3/2}\right)}{27 Q^4}.\label{eq:mub1}
\end{align}
This condition also constrains the maximum value of the electric charge, $Q \leq \sqrt{4/3} M$, in order for $\mu$ to have a real value. When $Q = 0$, the range of $\mu$ becomes $\mu < 1/16$, recovering the $\mu$ bound in \cite{Nielsen:2017hxt}, and satisfies the MTS stability condition in (\ref{eq:mubound}). For the saturated case $Q = \sqrt{4/3} M$ the bound on $\mu$ becomes $\mu < 1/12$. Note that the existence of the inner and outer horizons requires $Q < M$ (from (\ref{eq:rpm})) so that $Q \leq \sqrt{4/3} M$ always holds. Finally, in the case that $Q = M$, the bound on $\mu$ is $\mu < 2/27$. These various bounds on $\mu$ here, allowing for the existence of the conformal Killing horizons $r_i$, are well within the bound \eqref{eq:mubound} as is expected.

In fact, if neither $Q=M$ nor $Q=0$, it is not a very difficult exercise to check that the inequality \eqref{eq:mub1} imposes a bound on the mass-charge ratio:
\begin{align}
\frac{Q^2}{M^2}\lessapprox \frac{1}{25},\label{eq:mub2}
\end{align}
a sharp refinement of the bound $Q^2/M^2\leq4/3$.

Now, we return to the conformal Killing horizons located at $r_i$, $i\in\{1,2,3\}$. Let us expand around small $\mu$: 
\begin{align}
r_1 &\sim \frac{M}{2 \mu }+\frac{1}{2} (-4 M+v-v_0) +2 \mu  \left(\frac{Q^2}{M}-4 M-v+v_0\right) + \cdots, \\
r_2 &\sim M-\sqrt{M^2-Q^2} + \mu  \left(\frac{\left(\sqrt{M^2-Q^2}-M\right) \left(\sqrt{M^2-Q^2}-3 M-v+v_0\right)}{M}-\frac{Q^2}{\sqrt{M^2-Q^2}}\right) + \cdots, \\
r_3 &\sim M+\sqrt{M^2-Q^2} + \mu  \left(\frac{\left(\sqrt{M^2-Q^2}+M\right) \left(\sqrt{M^2-Q^2}+3 M+v-v_0\right)}{M}+\frac{Q^2}{\sqrt{M^2-Q^2}}\right) + \cdots .
\end{align}
We can then easily check that all three horizons are located at positive values of $r$. (The authors in \cite{tarafdar2023slowly} established that the \textit{slowly evolving horizons} (SENS), a weakly perturbed isolated horizon, coincides with the outer conformal Killing horizon.) Along the $r$ axis, the conformal Killing vector $\xi^a$ becomes spacelike for the regions $r > r_1$ and $r_2 < r < r_3$, timelike for the regions $r_3 < r < r_1$ and $r < r_2$, and null at $r=r_i$. (Notice that in the region $r_2 < r < r_-$, the conformal observers are not defined, which is in contrast to the timelike observers moving with $v$-time, which is defined for $r<r_-$.) In the static limit $\mu \rightarrow 0$, \(r_2\) and  \(r_3\) simply correspond to the inner and outer Killing horizon in Reissner-Nordst\"{o}m spacetime respectively, while $r_1$ blows up to infinity. 

Let us now check the energy condition (\ref{eq:ECdn}). The null energy condition is satisfied in regions $r_{\textrm{E}}$ where
\begin{align}
r_{\textrm{E}} \geq \frac{Q^2 (M+\mu  (v-v_0))}{M^2}.
\end{align}
Due to the small $\mu$ expansion, we can clearly make a comparison for the location of horizons as follows 
\begin{align}
r_1>r_3>r_+> r_{\textrm{E}} > r_->r_2,
\end{align}  
confirming the intuition that the outer $r_+$ (and consequently the inner $r_-$) horizon lies behind the $r_3$ conformal Killing horizon. Thus, the null energy condition is satisfied for the outside of the outer horizon $r_+$, but violated somewhere between the outer $r_+$ and inner $r_-$ horizons. This might indicate that the linear charged Viadya spacetime is unphysical. To resolve this problem, there have been studies to remove the unphysical region and to glue another spacetime \cite{AOri_1991,Booth:2015kxa}. 

\subsection{Surface gravity and thermodynamics}

We now compute the surface gravities for the conformal Killing horizons $r_i$.

On the conformal Killing horizons, the surface gravity can be defined in the following three ways \cite{Jacobson:1993pf}:
\begin{align}
\nabla_{a} (\xi^b \xi_b) &= - 2 \kappa_1 \xi_a, \label{eq:kappa1}\\
\xi^{b} \nabla_{b} \xi^{a} &= \kappa_2 \xi^{a}, \\
(\kappa_3)^2 &= - \frac{1}{2} (\nabla^a \xi^{b})  (\nabla_{[a} \xi_{b]}). \label{eq:kappa3}
\end{align}
These definitions of $\kappa_{i}$ all coincide for Killing vector case, but for conformal Killing vectors, they are related as
\begin{align}
\kappa_1 = \kappa_2 - 2 \varphi = \kappa_3 - \varphi. 
\end{align}

Applying the Lie derivative $\mathcal{L}_{\xi}$ to (\ref{eq:kappa3}), and using the relation
\begin{align}
\nabla_{c}\nabla_{a} \xi_{b} = \xi^{d}R_{dcab} + g_{ab} \nabla_{c} \varphi + g_{bc} \nabla_{a} \varphi - g_{ac} \nabla_{b} \varphi,
\end{align}
yields 
\begin{align}
2 \kappa_3 \mathcal{L}_{\xi}\kappa_1 = 0, 
\end{align}
which implies that $\kappa_1$ is constant along each of the null CKV curve. Moreover, $\kappa_1$ is conformally invariant and so remains the same even if Killing vectors defined in (\ref{eq:CKnK}) apply,
\begin{align}
\bar{\nabla}_{a} (\bar{\xi}^b \bar{\xi}_b) = - 2 \kappa_1 \bar{\xi}_a.
\end{align}
Thus, the surface gravity $\kappa_1$ is the one that is identified to Hawking temperature in a static spacetime as follows
\begin{align}
T= \frac{\kappa_1}{2 \pi}. \label{eq:HKTkappa}
\end{align}
In an asymptotically flat spacetime, Hawking temperature is defined by using the timelike Killing vector which is normalized to be unity at infinity. Here our dynamical spacetime is asymptotically flat, but the norm of CKV diverges at  infinity. This fact made us to use the CKV that is normalized to unity when $\mu \rightarrow 0$ at infinity.  The same normalization is done for a static spacetime. 

Calculating $\kappa_1$ via (\ref{eq:kappa1}) and expanding them around small $\mu$, it yields
\begin{align}
\kappa_1|_{r_1} =&  - \frac{1}{M} \mu + \frac{4}{M} \mu^2 + \cdots, \label{eq:r1}\\
\kappa_1|_{r_2(r_3)} =& \frac{-2 M^3+2 M Q^2 \mp \left(2 M^2-Q^2\right) \sqrt{M^2-Q^2}}{Q^4} \nonumber\\
&+ \frac{\mp M-4 \sqrt{M^2-Q^2}}{M \sqrt{M^2-Q^2}} \mu - \left(\frac{2}{M} \mp \frac{4 M^4- 6 M^2 Q^2 + 3 Q^4}{2 M^2 \left(M^2-Q^2\right)^{3/2}} \right) \mu^2 + \cdots , \label{eq:r23}
\end{align}
where $\kappa_1|_{r_2}$ takes the upper `$-$' sign and $\kappa_1|_{r_3}$ takes the lower `$+$' sign in (\ref{eq:r23}).  We also can directly compute the Hawking temperature from the static metric. Let us employ the metric (\ref{eq:metricC21}) and transform it to $(T,R)$ coordinate via 
\begin{align}
V = T + R_{*}, \qquad R_{*} = \int \frac{1}{\tilde{F}(R)} \rmd R.
\end{align}
This yields the following metric and Killing vector as follows:
\begin{align}
\rmd s^2 &= - \tilde{F}(R) \rmd T^2 + \frac{1}{\tilde{F}(R)} \rmd R^2 + R^2 \rmd \Omega_2, \\
\xi^{a}&= \{1,0,0,0 \}.
\end{align}
This static spacetime possesses three horizons at $\tilde{F}(R)=0$, which are related to the conformal Killing horizons 
\begin{align}
R_1 = \frac{M}{m(v)}r_1, \qquad R_2 = \frac{M}{m(v)}r_2, \qquad R_3 = \frac{M}{m(v)}r_3,
\end{align}
and $R_{i} \approx r_{i}$ for a small $\mu$ limit since $m(v)=M$ there. The Hawking temperatures are computed as 
\begin{align}
T_i = \frac{\tilde{F}'(R)}{4 \pi} \bigg|_{R=R_i} = \frac{\kappa_i}{2 \pi}
\end{align}
and if we expand $T_i$ for small $\mu$, they agree with (\ref{eq:HKTkappa}), applying (\ref{eq:r1}) and (\ref{eq:r23}) .

\subsection{Case of $M=Q$}

In the extremal case $M=Q$, $r_+=r_-=r^{\ast}=m(v)$, and $\mathcal{L}_l\chi |_{r^{\ast}}=\mathcal{L}_n\chi |_{r^{\ast}}=0$. The null energy condition (marginally) holds but the extremal horizon at $r^{\ast}=m(v)$ encloses no trapped surface, even though $C = \mu > 0$. (The evolution away from the extremal case of the metric \eqref{gvc1}, has been considered in detail by Booth \cite{Booth:2015kxa}, where the forms of the mass and charge were kept general.)  However, the conformal Killing horizons do not degenerate for $M=Q$. The conformal Killing horizons are now located at
\begin{align}
r_1 &= \frac{M}{2 \mu } +\frac{1}{2} (-4 M+v-v_0) -2 \mu  (3 M+v-v_0) + \cdots, \\
r_2 &= M-\sqrt{2} M \sqrt{\mu}+ (3 M+v-v_0) \mu + \cdots,\\
r_3 &= M + \sqrt{2} M \sqrt{\mu}+ (3 M+v-v_0)\mu  + \cdots,
\end{align}
and hence, there is a separation $r_2-r_3\neq0$. The non-vanishing surface gravity is computed at each $r_i |_{M=Q}$ horizon as
\begin{align}
\kappa_1|_{r_1} = - \frac{\mu}{M} + \cdots, \qquad \kappa_1|_{r_2} = - \frac{\sqrt{2} \sqrt{\mu} + 4 \mu}{M} + \cdots, \qquad  \kappa_1|_{r_3} = \frac{\sqrt{2} \sqrt{\mu }-4 \mu }{M} + \cdots.
\end{align}

As stated in the previous section, the conformal Killing horizons in the dynamical spacetime correspond to the horizons in static spacetime, under the conformal map, which are generated when the $g_{VV}$ component of the metric function in (\ref{eq:conformalmap}) vanishes. The surface gravities of these horizons yield the non-zero Hawking temperature in static spacetime. Therefore, even in the degenerate case for the linear charged Vaidya metric, where the $r_{\pm}$ horizons coincide, the Hawking temperature is non-zero in the associated static spacetime, since the conformal Killing horizons do not coincide. 

\section{Conclusion}

We have considered a dynamical spacetime in asymptotically flat spacetime, specifically the charged Vaidya spacetime in the ingoing Eddington-Finkelstein coordinate \cite{Vaidya:1970v1}. This spacetime describes the absorption of matter fields into black holes. We first found that the charged Vaidya spacetime admits a conformal Killing vector $\xi^a$ when the electric charge function $q(v)$ is proportional to the mass function $m(v)$ and the mass function is linear in $v$. This conformal Killing vector field appears to be the unique (modulo constant scaling) conformal Killing vector, admitted by the `linear' charged Vaidya metric. We accordingly demonstrated the conformal Killing vector found is indeed a homothetic Killing vector $4\varphi=\nabla_a\xi^a=\mbox{const.}$ and not a proper one, i.e. a non-constant $\varphi$. This induces the conformal map via which we obtained static spacetimes that can take various forms depending on the conformal factor $\Omega(v,r)$ that we choose. Making the choice $\Omega = m(v)/M$, we obtained the charged black hole solutions with a cosmological horizon. However, this spacetime is neither asymptotically flat nor asymptotically de Sitter. This solution is known in the literature as the black hole spacetime surrounded by quintessential matter fields in \cite{Kiselev:2002dx} or the solution that are obtained by employing anisotropic matter fields and Maxwell field in \cite{Jeong:2023hom}. 

Secondly, we investigated the properties of the various horizons. The linear charged Vaidya spacetime is known to admit an inner ($r_-$) and an outer ($r_+$) horizon  where the $g_{vv}$ component of the metric vanishes. Using the characterization via the expansion of the null generators, it is easily verified that the outer horizon is a dynamical horizon with trapped surfaces just to the inside. The appropriate upper bound on mass density $\mu$ of the null dust, allowing for the existence of trapped surfaces just to the inside of the dynamical horizon, is established.

The charged linear Vaidya spacetime also admits conformal Killing horizons, in addition to the two $r_{\pm}$ horizons, where the norm of conformal Killing vector vanishes. In particular, three conformal Killing horizons are generated, which we identified as the cosmological conformal Killing horizon ($r_1$), the inner conformal Killing horizon ($r_2$), and the outer conformal Killing horizon ($r_3$). The bound on $\mu$ for the existence of the conformal Killing horizons is found to be sharper than the case for the inner and outer $r_{\pm}$ horizons. This provides us with a required limit on the charge-mass ratio: $Q^2/M^2\lessapprox 1/25$. In the static limit $\mu \rightarrow 0$, the linear charged Vaidya spacetime approaches the Reissner–Nordström metric. The cosmological horizon expands to infinity, and from the inner and outer conformal Killing horizons $r_2$ and $r_3$, we respectively recover the standard inner and outer horizons in the Reissner–Nordström metric.

The null energy condition also imposes an order on the locations of the horizons. Indeed, the cosmological conformal Killing horizon is at the higher end of this hierarchy. We also have that $r_3>r_+$, which is as expected. Crucially, $r_1>r_2$, i.e. the inner conformal Killing horizon lies to the `inside' (a timelike region) of the inner $r_-$ horizon, where the energy condition fails. (The null energy condition holds for $r>r_+$.)

The surface gravity of the conformal Killing horizons were also analyzed. Surface gravity can be defined in three ways $\kappa_1,\kappa_2,$ and $\kappa_3$ \cite{Jacobson:1993pf}, with $\kappa_1$ obtained from the gradient of the norm of the conformal Killing vector. For Killing vectors, all three definitions agree. However, for the conformal Killing case, the three definitions differ by some scale of the factor $\varphi$. We computed the surface gravity $\kappa_1$, which is conformally invariant and constant along the conformal Killing trajectory. This definition is identified with the Hawking temperature in the static spacetime. Since the norm of the conformal Killing vectors diverges as $r$ goes to infinity, we required it to be unity when $\mu = 0$. Therefore, the Hawking temperature in this paper is constructed by using the (conformal) Killing vectors normalized in this manner. Interestingly, the conformal Killing horizons in the dynamical spacetime map to horizons in the corresponding static spacetime that are generated when $g_{VV}= 0$, with $V$ being the `time' coordinate in the static spacetime. This allowed us to confirm that the surface gravity that we chose indeed corresponds to the Hawking temperature.

Finally, we examined the case where the mass $M$ is equal to the electric charge $Q$. This leads to the coincidence of the inner and outer $r_{\pm}$ horizons at $r=m(v)$ in the linear charged Vaidya spacetime. While the expansion parameter characterizing the horizon $C=\mu$ is non-negative, this extremal horizon encloses no trapped surface. Unlike the $r_{\pm}$ horizons, the conformal Killing horizons do not coincide at $M=Q$, and still induce surface gravity at each surface. Therefore, the degenerate case does not yield vanishing surface gravity in the linear charged Vaidya spacetime, and hence a non-zero Hawking temperature in the static spacetime.

\section*{Acknowledgments}

We wish to thank Ivan Booth for some useful discussions. M.P was supported by the Institute for Basic Science (Grant No. IBS-R018-Y1). S.K was supported by the Basic Science Research Program through the National Research Foundation of Korea (NRF) funded by the Ministry of education (Grant No. NRF-2021R1A2C1005748). A.M.S. was supported by the Basic Science Research Program through the National Research Foundation of Korea (NRF) funded by the Ministry of education (Grant No. NRF-2021R1A2C1005748) and (Grant No. NRF-2022R1I1A1A01053784). We appreciate APCTP and the Centro de Ciencias de Benasque Pedro Pascual for its hospitality during the completion of this work.

\bibliographystyle{ieeetr}
\bibliography{CKVbib}

\end{document}